\documentclass[a4paper]{article}

\usepackage[T1]{fontenc} 
\usepackage{lmodern}  
\usepackage{fancyhdr} 
\usepackage[utf8]{inputenc} 
\usepackage{hyperref} 
\usepackage{listings} 
\usepackage{amssymb}  
\usepackage{courier}  
\usepackage{graphicx} 
\usepackage[numbers]{natbib} 
\usepackage[linesnumbered,vlined,boxed]{algorithm2e} 
\usepackage[top=3cm,bottom=3cm,left=3cm,right=2cm]{geometry} 
\newenvironment{buit} 
  {\setlength{\parindent}{10pt}
   \itemize      
   }        
  {\enditemize
   \vspace*{\baselineskip}
   \setlength{\parindent}{0pt}}

\newtheorem{prop}{Property}

\title{Statistical algorithms for particle trajectography}
\author{Frédéric Magniette\\
\footnotesize Laboratoire Leprince-Ringuet (LLR), \'Ecole polytechnique, CNRS/IN2P3, F-91128 Palaiseau, France}
\begin{document}
\maketitle

\begin{abstract}
The various algorithms used to extrapolate particle trajectories from measurements are often very time-consuming with computational complexities which are typically quadratic. In this article, we propose a new algorithm called GEM with a lower complexity and reasonable performance on linear tracks. It is an extension of the EM algorithm used to fit Gaussian mixtures. It works in arbitrary dimension and with an arbitrary number of simultaneous particles. In a second part, we extend it to circular tracks (for charged particles) and even a mix of linear and circular tracks. This algorithm is implemented in an open-source library called ``libgem'' and two applications are proposed, based on data-sets from two kind of particle trackers.
\end{abstract}

\section{Introduction}

Particle detectors rely on devices, called trackers, recording points where particles passed through them and allowing thus to infer the parameters of their movements. This operation relies on the estimation of the trajectory of the particles from the measured points. As the detector is often in a magnetic field, the trajectories are linear for neutral particles or helical for charged ones. In the latter case, the radius of the trajectory is proportional to the momentum of the particle.

Extrapolating linear trajectories is done easily by a linear regression if the detector manages only one particle at a time. But as the phenomena we look for are extremely rare, physicists designing particle accelerators tend to increase the rate of event pile-up (i.e. simultaneous events) to aquire more data and get more chance to observe these events. Thus, in such detectors, it is necessary to use algorithms, which are able to discriminate the different tracks and reconstruct them at the same time.

The Hough transform is a powerfull tool to make such reconstructions \cite{duda_hough}. In two dimensions, it is a spectral method based on a histogram of all the directions of the point pairs. By extracting the peaks of the histograms, the track directions can be reconstructed. This method allows also to make circular reconstruction through a generalized model. Unfortunately, the Hough transform is very CPU-consuming as stated in \cite{hough_survey}. The linear fit algorithm is typically in $O(n^2)$ in 2-dimensions (where $n$ is the number of points). When moving to 3-dimensions or to circular fit, the storage requirements and computational complexity increases a lot and can become intractable on online infrastructure, which are lightweight by construction. Other step by step methods, like Kalman filter \cite{kalman} or cellular automaton, can compute such fits, building a neighborhood and fitting linearly inside it. Their complexity is also very high and their performance can be incompatible with online treatment.

As the particle detectors raise-up in event frequency, triggering becomes more crucial. It becomes impossible to store all data produced by the detector and thus a choice has to be done. This step has to discriminate between interesting collisions and the others. Of course, online trajectography is a major tool in such triggers. It is then very interesting to have powerful fit algorithms with a complexity that is lower than quadratic.

Such algorithms exist, based on mixtures of Gaussian models. They are called EM algorithms (for Expectation-Maximization). They allow to separate Gaussian components from a signal. A linear extension has been proposed in \cite{turner2000} that allows to fit multiple tracks with a mixture of linear regressions. This algorithm has some limitations, incompatible with particle detector requirements: anisotropy, limitation to two dimensions and fixed number of tracks.

In this article, we propose a new version of the EM algorithm which solves all these problems. We obtain an adaptive fit (in number of tracks), fully isotropic and in arbitrary dimension. This algorithm has been implemented in a library coded in C for performance, and called ``libgem''. It has been tested on two data-sets coming from a time projecting chamber and a silicon tracker. We also propose an extension of the algorithm to handle circular tracks and we show that the algorithm is generic and can be extended to multiple other shapes. Finally, we propose another algorithm to combine the linear and circular fits and we test it on simulations.

\section{Background}

\subsection{Fit}
A fit is a procedure for evaluating the parameters of a model, driven by a set of data. This evaluation is done by using dedicated algorithms, named estimators, for each parameter. The quality of the fit can be evaluated by calculating an error function, usually a distance between each data point and the model. The best fit (generally just called the fit) is a fit minimizing the error function. If the fit is weighted, a weight noted $\tau_i$ is given to each point $p_i$ and the error function has to be weighted by the $\tau_i$ as well. If some distance function $d$ exists between the data points $p_i$ and the model $M$ of parameter $\theta$, the error function usually looks like

\begin{equation}
\varepsilon=\frac{\sum_{i=1}^n\tau_i d(p_i,M(\theta))}{\sum_{i=1}^n\tau_i}.
\end{equation}

\subsection{Gaussian mixture model}

The Gaussian mixture models are functions like

\begin{equation}
G(x,\Theta) = \sum_{k=1}^{K} \pi_k g(x,\theta_k),
\end{equation}

where $g$ is the Gaussian function of parameters $\theta_k$. The $\pi_k$ are the coefficients representing the normalized weighting of the mixture. $\Theta$ is a compact notation for all the parameters $\theta_k$ and $\pi_k$. Given the mixture, the hard part is to extract its parameters. This is the purpose of the EM algorithm, introduced in 1977 by Dempster {\em et al} \cite{dempster77}. EM stands for Expectation-Maximization which are the two phases of this iterative algorithm.

\begin{figure}[ht!]
\centering
\includegraphics[height=.30\textwidth]{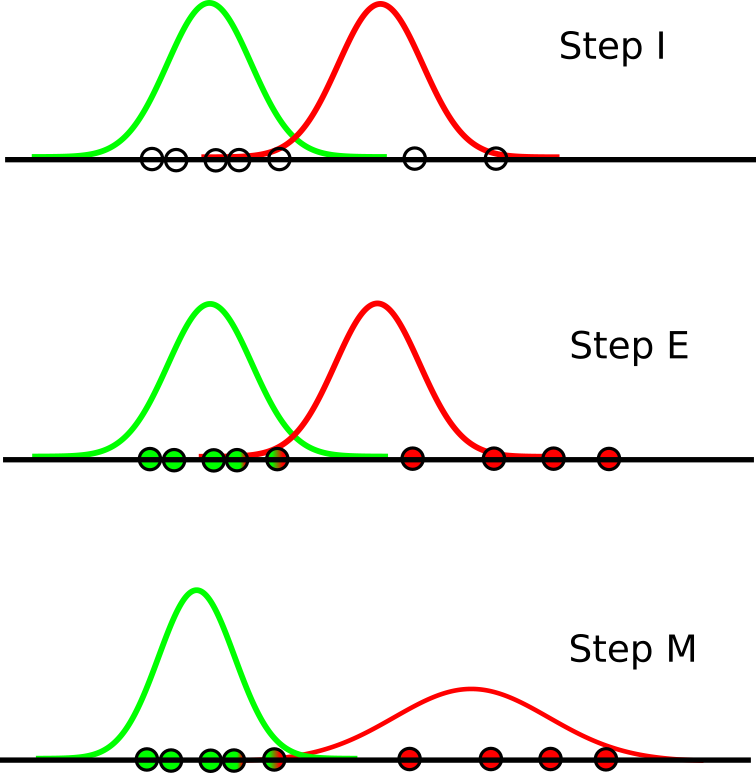}
\caption{The 3 steps of the EM algorithm: Initialization: the Gaussians are initialized randomly, Expectation: a probability of membership is computed for every point and every Gaussian, Maximization: the parameters of the gaussian are evaluated based on the point weights.}
\label{fig:em}
\end{figure}

Given a set of $n$ numbers $x_i$, we want to explain the distribution by a mixture of $K$ Gaussians. As illustrated in Figure \ref{fig:em}, we initialize the $K$ distributions  $N(\bar{x}_k,\sigma_k^2)$ randomly and then we apply the two following steps: 
\begin{buit}
\item Expectation phase: for each point $x_i$, we calculate the probability of membership for each Gaussian $\tau_{ik}$, based on the normalized likelihood function of the Gaussian distribution defined by
  \begin{equation}
    l_{ik}=\frac{1}{\sigma_k\sqrt{2\pi}} e^{-\frac{(x_i-\bar{x}_k)^2}{2\sigma_k^2}}.
  \end{equation}
  These coefficients are then normalized to obtain the $\tau_{ik}$.
\item Maximisation phase: for each Gaussian, we calculate the parameters based on the points weighted by the $\tau_{ik}$
  \begin{equation}
    \bar{x}_k^*=\frac{ \sum_{i=1}^{n} \tau_{ik}x_i }{ \sum_{i=1}^{n} \tau_{ik}} \;\;,\;\; s_k^*=\sqrt{ \frac{ \sum_{i=1}^{n} \tau_{ik}(x_i-\bar{x}_k^*)^2} {\sum_{i=1}^{n} \tau_{ik}} }.
  \end{equation}
\end{buit}

This algorithm is controlled by the log-likelihood given by $l=\log ( \prod_{i=1}^{n} P(x_i) )$. When $l$ becomes stable enough, the algorithm is converged. This convergence corresponds to a minimum of the log-likelihood. This minimum can be local, depending on the initialization. In some bad cases, the convergence can be very slow and the algorithm is considered as being divergent. To avoid these initialization problems, one can use meta-algorithms like \cite{smallem} for instance, which create a set of initial conditions and execute the algorithm on all of them to choose the best initialization point.

Another drawback of the EM algorithm is that the number of components is a parameter of the algorithm. In case this value is not known, again a meta-algorithm is necessary: different values have to be tested and the best solution is taken. For example, the BIC criterion published in \cite{schwartz78}, proposes a penalization of any model by a function of both the number of observations and the number of parameters.

\subsection{Linear regression mixture}

The EM algorithm has been adapted to linear regression mixture in \cite{turner2000}. Instead of extracting a Gaussian signal from a mixture, linear regressions surrounded by a Gaussian noise are extracted. For convenience, let's call this model LRMM (Linear Regression Mixture Model) in the sequel. Its general form for the $k^{th}$ component of the mixture is $y=x\beta_k+\varepsilon_k$. The error distribution $\varepsilon_k$ follows a centered Gaussian $N(0,\sigma_k^2)$.

The likelihood is not computable analytically, thus it is necessary to use the EM algorithm to estimate the mixture parameters. During the Maximization phase, the linear parameters and the Gaussian parameters are estimated sequentially. This method has been implemented in the ``mixreg'' package, described in \cite{mixreg} and usable in the statistical framework R. 

\section{Data description}

\begin{figure}[ht!]
  \centering
  \includegraphics[width=.32\textwidth]{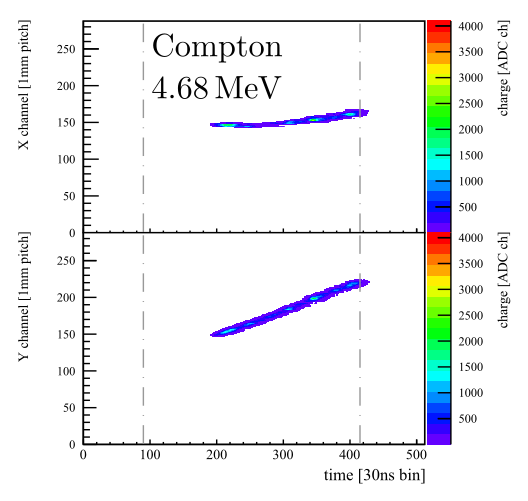}
  \hfill
  \includegraphics[width=.32\textwidth]{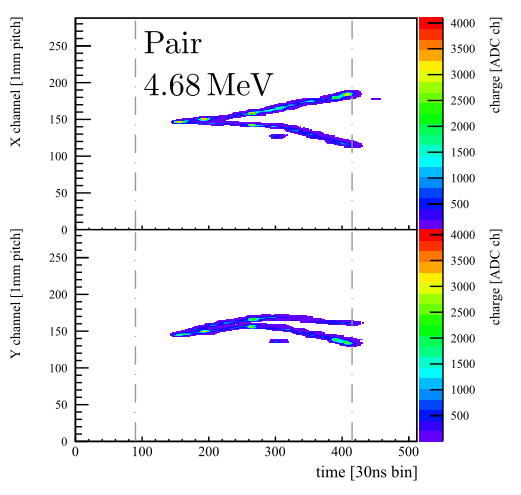}
  \hfill
  \includegraphics[width=.32\textwidth]{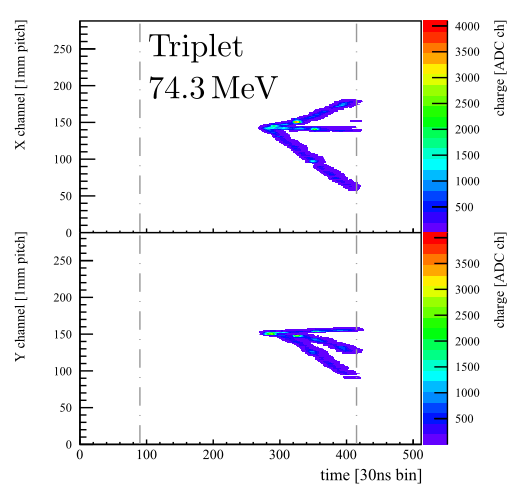}
  \caption{The three kind of tracks $x(t),y(t)$ generated by Harpo. On the left, a Compton scattering, in the middle, a pair production and, on the right, a triplet production. These tracks are perturbed by multiple scattering and electronic noise.}
  \label{fig:harpo_tracks}
 \end{figure}

Two different data sets are used for this study. One comes from a test-beam performed at NewSUBARU on a Time Projection Chamber (TPC), a gazeous tracker named Harpo and developed at LLR. The results of the test are detailed in \cite{HARPO17}. Harpo is developed to become a spatial gamma ray telescope and polarimeter. It is a drift chamber of 30 cm$^3$ filled with a mixture of Argon and Isobutane (95:5) at 1 bar. When an incident gamma crosses the gas, ionization electrons are produced. An electric field induces a drift of these electrons at 3.3 cm$/\mu$s and two perpendicular strips read them using gazeous electron multipliers (GEM \cite{gem} and Micromegas\cite{micromegas}). The corresponding data consist of coupled 2D tracks representing the two axes as a function of time $(x(t),y(t))$. These tracks, as shown on Figure \ref{fig:harpo_tracks}, can be composed of a single line corresponding to a Compton scattering ($e^- \gamma \rightarrow e^- \gamma$) , two lines depicting a pair production ($\gamma Z \rightarrow e^+ e^- Z$) or three lines corresponding to a triplet production ($\gamma e^- \rightarrow e^+ e^- e^-$). All these shapes are perturbed by the multiple scattering.

 The detector generates a big amount of points, typically too many for getting an adequate performance using reconstruction algorithms. Thus, we are using a pre-processing step to reduce the number of points. Applying a grid on the data, we take the barycenter of the points in any grid pixel. By this way, we get a parametrized reduction of the number of points. We can see on Figure \ref{fig:harpo_tracks} (in the middle at abscissa 300), that some stain can appear, artefacts of the reading system. We remove them during the pre-processing by  calculating the connected components and removing those which size is lower than the parametrized threshold.

\begin{figure}[ht!]
  \centering
  \includegraphics[width=.45\textwidth]{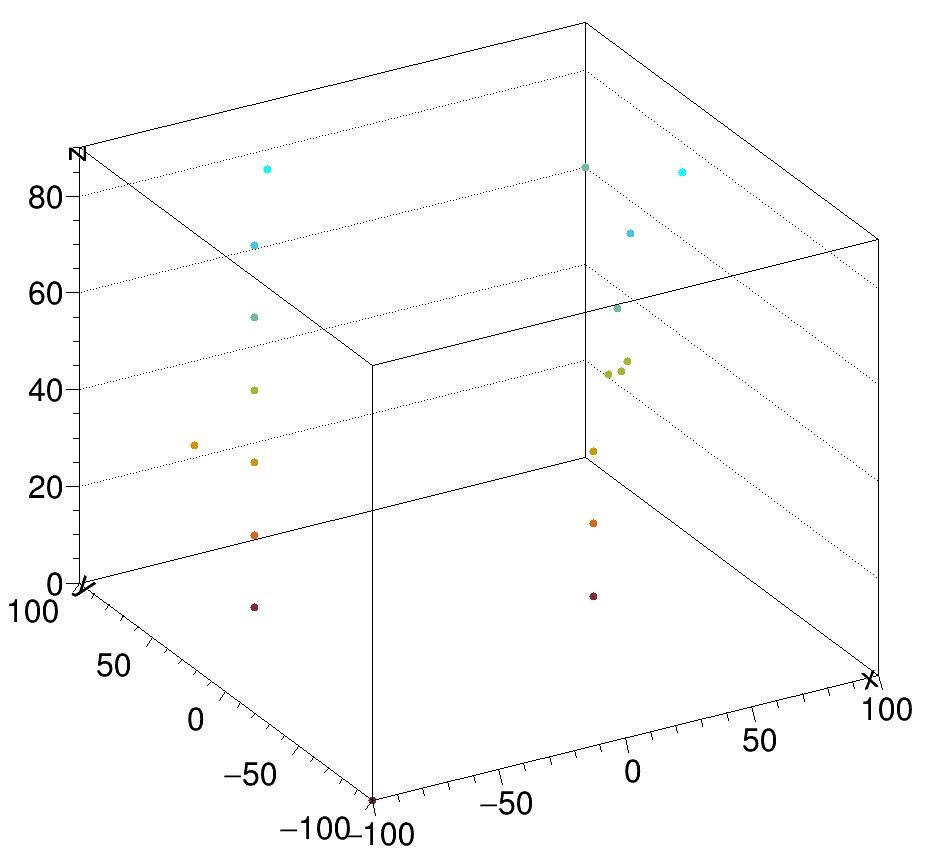}
  \caption{Typical pile-up track of SiW-Ecal in tracker mode: two electrons hit the detector during the same integration period. Their tracks are almost linear, with some extra points due to pair production in the reading system.}
  \label{fig:ecal_3D}
\end{figure}
 
The second data set comes from a test-beam of the SiW-Ecal ILD prototype described in \cite{balag}. This electromagnetic calorimeter developed at LLR, is made of an alternance of tungsten absorbers and silicon pixels. During the test-beam at DESY in 2017, the single particule response (MIP) has been calibrated without tungsten absorbers as explained in \cite{tbres}. In that case, the electromagnetic calorimeter acts as a silicon tracker. Detecting angular deviations of the tracks is an important goal. Indeed, the energy deposit is proportional to the length of material crossed by the particle. Thus, the MIP calibration depends on this angle. As represented on Figure \ref{fig:ecal_3D}, its tracks are 3-dimensional but made of few layers (presently 7).

In about 10\% of cases, pile-up appears, as on Figure \ref{fig:ecal_3D} with two tracks or rarely with three. Thus it is necessary to have an adaptive algorithm to fit properly these data.

\section{Anisotropy of LRMM}

\begin{figure}[h!]
  \centering
  \includegraphics[width=.42\textwidth]{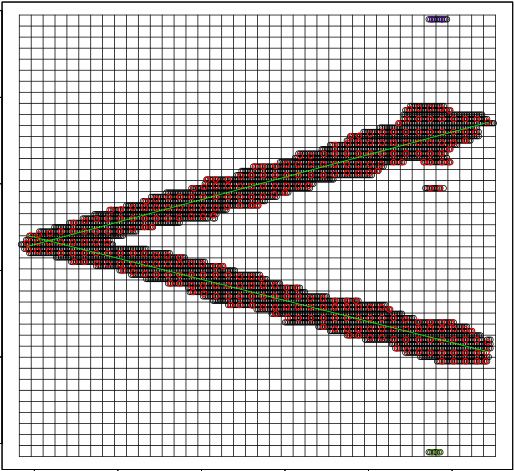}
  \hfill
  \includegraphics[width=.42\textwidth]{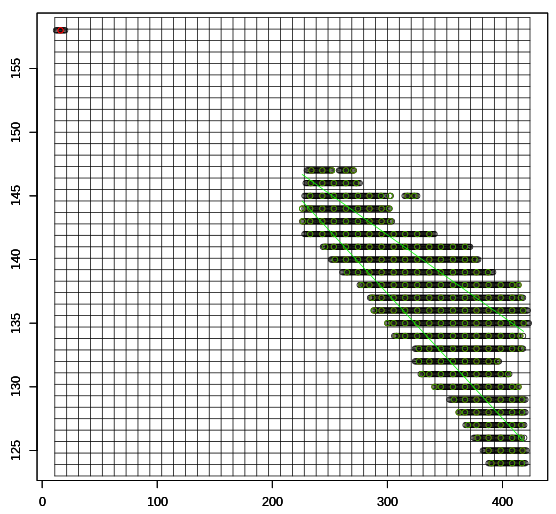}
  \caption{LRMM fits on Harpo data with the ``mixreg'' package. In both cases, the represented interaction is a pair production. LRMM fits well the data.}
  \label{fig:harpo_fit}
\end{figure}

LRMM gives good results on Harpo pre-processed data, as represented on Figure \ref{fig:harpo_fit}.

On the left, we see a pair track. The original points (measured by the TPC) are in black. The red points are the result of the barycentric pre-processing induced by the black grid. The small stains on the right are removed and the fit is the pair green lines. We see that the lines fit correctly the data and gives a credible interaction vertex. On the right, we can see another pair with a smaller angle. The LRMM algorithm gives a good approximation of the lines even in this entangled case.

\begin{figure}[ht!]
  \centering
  \includegraphics[width=.35\textwidth]{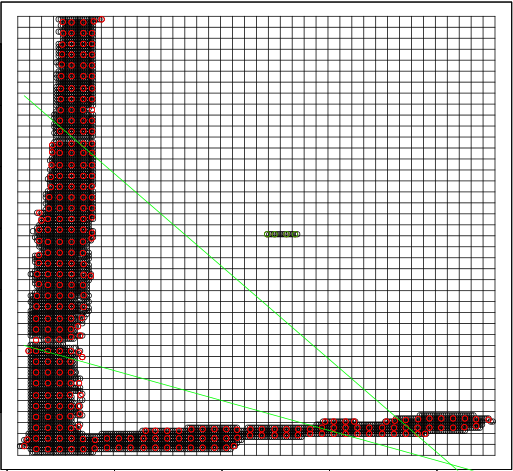}
  \hfill
  \includegraphics[width=.40\textwidth]{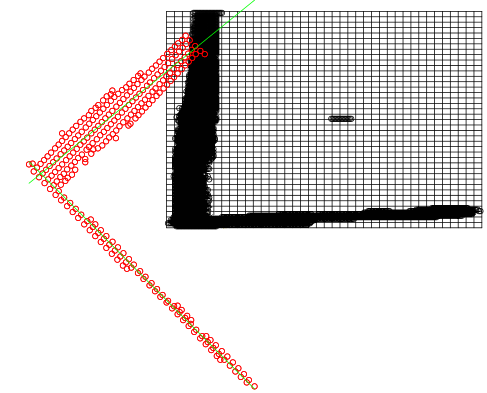}
  \caption{If a vertical line has to be fitted, LRMM diverges and gives bad results by anisotropy of the algorithm. Indeed, if the data are rotated, the fit becomes correct again.}
  \label{fig:harpo_badfit}
\end{figure} 

In a small but not negligible fraction of the cases, the algorithm diverges, as shown on the left of Figure \ref{fig:harpo_badfit}. The fit, represented by the two green lines, does not correspond to the data. This divergence appears only when a vertical line has to be fitted, indicating an anisotropy in the algorithm. Indeed, if the same data is fitted after being rotated by $\pi/8$ radians as shown on the right of Figure \ref{fig:harpo_badfit}, the algorithm converges.

This anisotropy comes from the estimator of the Gaussian error variance defined by 
\begin{equation}
\sigma_k^2=\frac{\sum_{i=1}^{n} \tau_{ik}(y_i-x_i\beta_k)^2}{\sum_{i=1}^{n} \tau_{ik}}.
\end{equation}

\begin{figure}[ht!]
  \centering
  \includegraphics[width=.4\textwidth]{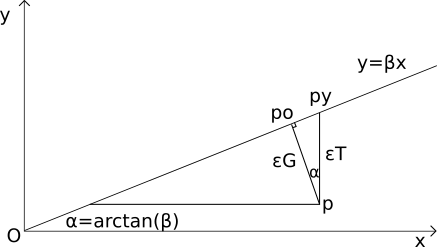}
  \caption{Linear model error $\varepsilon T$ vs orthogonal distance $\varepsilon G$. The distance used in the linear model is not invariant by frame rotation.}
  \label{fig:gem_erreur}
\end{figure}

The error term $(y_i-x_i\beta_k)$ is represented on Figure \ref{fig:gem_erreur} as $\varepsilon T$, the length of the segment p-py. It means that the error is calculated on the basis of a projection, parallel to the y axis. Obviously, if the frame $(O\overrightarrow{x}\overrightarrow{y})$ rotates clockwise, the $\varepsilon T$ value increases. This is the basic reason of the observed anisotropy. 

To eliminate it, we should replace this distance by an orthogonal distance, which does not vary with rotation, by construction. It is represented by $\varepsilon G$ on the Figure \ref{fig:gem_erreur} and it is the smallest distance between p and the line. As the p-po-py triangle is rectangle, we have $\cos(\alpha)=\frac{\varepsilon_G}{\varepsilon_T}=\cos(\arctan(\beta))$. As $\cos(\arctan(\beta))=\frac{1}{\sqrt{1+\beta^2}}$,
\begin{equation}
\varepsilon_G=\frac{\varepsilon_T}{\sqrt{1+\beta^2}}.
\end{equation}

This provides a new estimator for the $\varepsilon_k$ parameter that guarantees the isotropy of the algorithm. As $1/\sqrt{1+\beta^2}$ is continuous, we can apply the portmanteau theorem \cite{portemanteau}, which guarantees the consistency of the new estimator.

We can also extend the definition of the model by using a vector formulation of the orthogonal distance $d(P,(d))=\frac{\overrightarrow{RP} \cdot \overrightarrow{D}}{\mid\mid\overrightarrow{D}\mid\mid}$ \footnote{The double bar notation is the norm of the vector and $\cdot$ is the scalar product.}. Where the line $(d)$ is represented by a reference point $R$ and a director vector $\overrightarrow{D}$. Thus, we can generalize the algorithm to an arbitrary dimension if we can get a weighted linear regression in vectorial form.

\section{Arbitrary dimension weighted linear regression}

In this section, we describe an algorithm, called $\Phi_{line}$, that solves linearly the arbitrary dimension weighted linear regression. Its input is made of $n$ points named $P_i$, belonging to a space of dimension $Q$ and a vector of weights $\tau_i$. As the space dimension is arbitrary, it is convenient to represent a line by a reference point called $R$ and a director vector called $\overrightarrow{D}$. An obvious choice for $R$ is the weighted barycenter of the $P_i$ defined by

\begin{equation}
R=\frac{\sum_{i=1}^n\tau_iP_i}{\sum_{i=1}^n\tau_i}.
\end{equation}

To calculate $\overrightarrow{D}$, we can use the same technique as in Hough transform, building a weighted histogram and searching for a peak to find the direction. But as only one peak is expected and that the error is Gaussian, this peak coincides with the empirical mean of the directions. This leads to an analytical formulation 

\begin{equation}
\overrightarrow{D}=\sum_{i=1}^n\sum_{j=i+1}^n\tau_i\tau_j(P_i-P_j).
\end{equation}

As the norm of the director vector has no importance, we do not normalize the equation to simplify the notation. This is summed up on Algorithm \ref{alg:phi_line_exact}. Its complexity is $O(n^2)$. As this routine is the key part of the EM algorithm, it is called very often and this complexity can be a major drawback in case of big data-set. This is why we propose a heuristic called ``Refsplit'' to replace it.

\SetKwInOut{Variables}{Variables}
\begin{algorithm}[H]
  \Variables{$\varepsilon$ : array of orthogonal distance between the points and the fit line}
  $R=\frac{\sum_{i=1}^{n} \tau_i P_i}{\sum_{i=1}^{n} \tau_i}$\;
  $\overrightarrow{D}=\overrightarrow{0}$\;
  \For{$i \in 1..n$}{
    \For{$j \in i+1..n$}{
      $\overrightarrow{D}=\overrightarrow{D}+\tau_i\tau_j(P_i-P_j)$\;
    }
  }
  $\varepsilon_{i=1..n}=\mid\mid\overrightarrow{RP_i}-\frac{\overrightarrow{RP_i}.\overrightarrow{D}}{\mid\mid\overrightarrow{D}\mid\mid^2}\overrightarrow{D}\mid\mid$\;
  \Return $(R,\overrightarrow{D}),\varepsilon$\;
  \caption{$\Phi_{line}$ exact algorithm}
  \label{alg:phi_line_exact}
\end{algorithm}

The idea of the ``Refsplit'' heuristic is to replace the point pair vectors by the reference-point vectors. The direction of the director vector has to be the weighted mean direction of this collection of $\overrightarrow{RP_i}$ vectors. To find this direction, we could naively take the weighted mean of these vectors

\begin{equation}
\overrightarrow{D}=\sum_{i=1}^n\tau_i(P_i-R).
\end{equation}

However this sum is null by construction. Using the property that a vector has the same direction as its opposite, we invert some of these vectors to get a non null sum. The idea is to split the vectorial space in two sub-spaces separated by a hyperplane. All vectors in a part remain the same, and all the others are inverted. For example, we can divide the space between the vectors with $x \geq 0$ and the other with $x<0$. As we know our vectors sum is null, then the half of them are reversed, and the other half are not, compensating each others. Let's define a function $\delta(\overrightarrow{V})$ which returns $1$ if $V_x \geq 0$ and $-1$ if $V_x<0$. We can now define the vector director by

\begin{equation}
\overrightarrow{D}=\sum_{i=1}^n\tau_i\delta(\overrightarrow{RP_i})\overrightarrow{RP_i}.
\end{equation}

The sum is not null anymore and the vector $\overrightarrow{D}$ obtained is a good approximation of the one obtained by the exact method as soon as the axis is properly choosen. To do so, we compute an amplitude vector

\begin{equation}
\overrightarrow{A}=\sum_{i=1}^{n}\mid\overrightarrow{RP_i}\mid,
\end{equation}

where the $\mid.\mid$ denotes the absolute value, coordinate per coordinate. The preferred coordinate is the coordinate of $\overrightarrow{A}$ with the highest value. The ``Refsplit'' heuristic is decribed on Algorithm \ref{alg:phi_line_heur}. Its complexity is $O(n)$. 

\SetKwInOut{Variables}{Variables}
\begin{algorithm}[H]
  \Variables{a : preferred axis\\$\varepsilon$ : array of orthogonal distance between points and fit line\\$\delta$ : reversion indicator}
  $R=\frac{\sum_{i=1}^{n} \tau_i P_i}{\sum_{i=1}^{n} \tau_i}$\;
  $\overrightarrow{A}=\sum_{i=1}^{n}\mid\overrightarrow{RP_i}\mid$\;
  $a=\max_{i=1}^{Q}A_i$\;
  \For{$i \in 1..n$}{
    \lIf{$\overrightarrow{RP_{ia}}<0$}{$\delta_i=-1$}
    \lElse{$\delta_i=1$}
  }
  $\overrightarrow{D}=\frac{\sum_{i=1}^{n} \tau_i \delta_i \overrightarrow{RP_i}}{\sum_{i=1}^{n} \tau_i}$\;
  $\varepsilon_{i=1..n}=\mid\mid\overrightarrow{RP_i}-\frac{\overrightarrow{RP_i}.\overrightarrow{D}}{\mid\mid\overrightarrow{D}\mid\mid^2}\overrightarrow{D}\mid\mid$\;
  \Return $(R,\overrightarrow{D}),\varepsilon$\;
  \caption{$\Phi_{line}$ algorithm based on the ``Refsplit'' heuristic}
  \label{alg:phi_line_heur}
\end{algorithm}

As can be seen, the algorithm $\Phi_{line}$ returns two results: the parameters of the line composed of $R$ and $\overrightarrow{D}$, and an array of orthogonal distances between the points and the line. This allows to group all geometry related operations to allow geometrical extension. 

A 2D simulation study has shown that the raw error induced by this heuristic does not exceed $7.10^{-6}$ on the ratio $D_x/D_y$ which characterizes completely the direction. The study includes varying number of points, exhaustive angle exploration by degree and dispersion variation (1620000 cases). The mean of the errors is around $10^{-11}$ and standard deviation is around $2.10^{-7}$. Such low values show that the ``Refsplit'' heuristic does not degrade the performance at all.

\section{Isotropic linear fit in arbitrary dimension}

In this section, we propose an extension of the EM algorithm, named ``GEM'' for generalized EM. It improves the LRMM algorithm by removing estimator anisotropy and by allowing to work in arbitrary dimension. 

The idea of the algorithm is based on four steps. First, we initialize the weights of the mixture by choosing uniform random and normalized values for the $\tau_{ij}$. Then GEM executes three steps in a loop:
\begin{buit} 
\item Regression: computation of the weighted fit by $\Phi_{line}$;
\item Maximization: computation of the Gaussian error parameters (based on the distance computed by $\Phi_{line}$);
\item Expectation: computation of the point weights (normalized likelihood of the Gaussian distribution).
\end{buit}

The parameters are $J$, the number of components in the mixture, $\lambda$, the Gaussian normalized likelihood and $l_{ij}$, the probability of membership of the point $i$ to the component $j$ of the mixture.

The convergence is decided on a criterion of proximity between the points and the fit. For each point $P_i$, we define $prox(P_i)$ its closest line. The convergence criterion is the sum of $d(P_i,prox(P_i))$ for all points. As soon as this criterion becomes stable, the algorithm is considered to have converged.

\SetKwInOut{Variables}{Variables}
\begin{algorithm}[H]
 \Variables{$L$ : array of lines parameters\\ $D$ : array of distances from point to line\\$\tau$ : matrix of weights}
 \For{$j \in 1..J$} {
   \For{$i \in 1..n$} {
     $\tau_{ij}=$ normalized random\;
   }
 }
 \Repeat{converged} {
   \tcc{Regression step}
   \For{$j \in 1..J$} {
     $L_j,D_j=\Phi_{line}(p,\tau_j)$\;
   }
   \tcc{Maximization step}
   \For{$j \in 1..J$} {
     $s_j^*=\sqrt{ \frac{ \sum_{i=1}^{n} \tau_{ij} D_{ij}^2} {\sum_{i=1}^{n} \tau_{ij}}}$\;
   }
   \tcc{Expectation step}
   \For{$ i \in 1..n$} {
     \For{$ j \in 1..J$} {
       $l_{ij}=\lambda(d_{ij},D_{j},\tau_{ij})$\; 
       $\pi_j=\frac{\sum_{i=1}^{n}{\tau_{ij}}}{n}$\;
       $\tau_{ij}=\frac{\pi_j l_{ij}}{\sum_{l=1}^{J}{\pi_l l_{il}}}$\;
     }
   }
 }
 \Return $L,\tau$
\caption{GEM Algorithm}
\label{alg:olrmm}
\end{algorithm}

\begin{figure}[ht!]
\centering
\includegraphics[width=.40\textwidth]{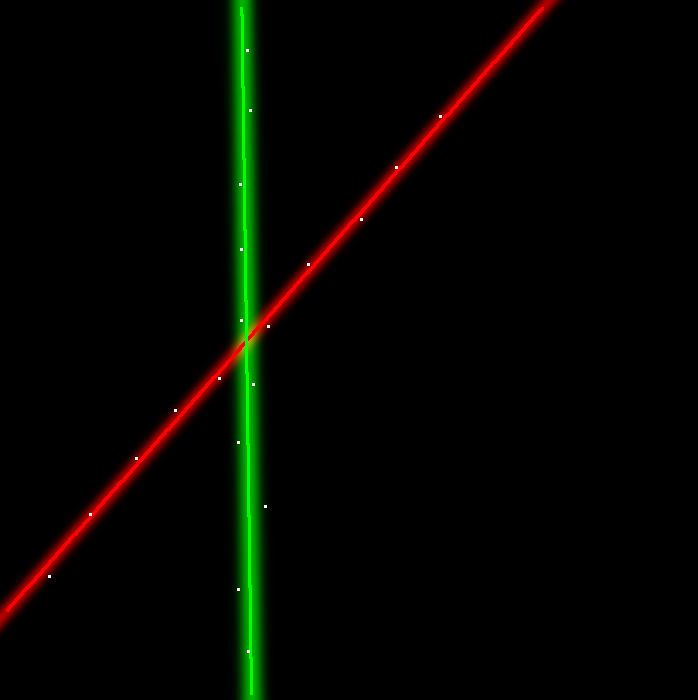}
\includegraphics[width=.40\textwidth]{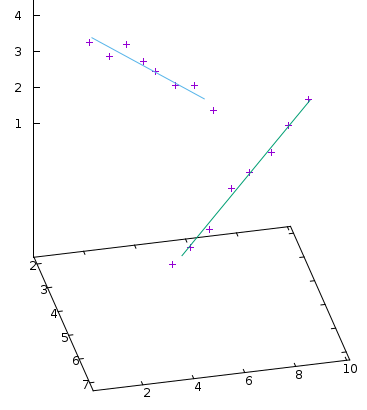}
\caption{GEM algorithm result in 2D and 3D on two noised lines}
\label{fig:olrmm}
\end{figure}

Figure \ref{fig:olrmm} shows the result of GEM on two sets of points. The first on the left is composed of two 2D-lines, slightly affected by a Gaussian noise. As one can see, the fit copes well with the data. Even if one of the two lines is vertical, the algorithm converges properly and the anisotropy has disappeared. On the right, the data set is composed of two lines in 3D. Again, the fit copes with data. 

As expected, we have solved the anisotropy and the dimension limitation of LRMM, but there are still two limitations left: the number of lines has to be given as a parameter and the algorithm is still very dependent on the initialization.

\section{Adaptive linear fit in arbitrary dimension}

In this section, we propose an algorithm that removes these two limitations at the same time. The idea is to begin with only one line and then to add new lines one by one until the fit becomes satisfying. It solves the initialization problem because, the new lines are not chosen randomly but to fit properly the worst fitted points. The question is how to build such a line and to decide when it is necessary to add one.

\newcommand\wfp{\mathit{wfp}}
\SetKwInOut{Parameters}{Parameters}
\SetKwInOut{Variables}{Variables}
\begin{algorithm}[H]
  \Parameters{$c$ : convergence criterion\\$s$ : scale criterion}
  \Variables{$D$ : array of parameters of shape S\\$G$ : array of Gaussian distributions\\$\tau$ : fit weights\\$J$ : number of fit lines\\$converged$ : Boolean indicator of convergence}
  $D_1=\Phi_S(P,[1/n,1/n,...,1/n])$\;
  $G_1=error(D_1,P,\tau)$\;
  $J=1$\;
  \Repeat{converged} {
    $\wfp,prox_{\wfp}=worst\_fitted(P_i,D)$\;
    $J=J+1$\;
    \tcc{Weights for new line}
    \For{$P_i \in P$}{
      \If{$d(\wfp,p)<prox_{\wfp}$} {$\tau_{iJ}=1-10^{-5}$\\
        \For{$j \neq J$}{$\tau_{ij}=10^{-5}/(J-1)$}
      }
      \Else{$\tau_{iJ}=10^{-5}$\\
        \For{$j \neq J$}{$\tau_{ij}=\tau_{ij}-10^{-5}/(J-1)$}
      }
    }
    \tcc{Apply GEM and determine convergence}
    $D,G,\tau=GEM(J,P,\Phi_S)$\;
    $converged=True$\;
    \For{$ j \in 1..J$} {
      \If{$G_j.\sigma>s/D$}{$converged=False$}
    }
  }
  \Return $D,G,\tau,J$\;
\caption{MFit Algorithm}
\label{alg:mfit}
\end{algorithm} 

To choose the new line, we search the worst fitted point, the one which has the highest value of $d(P_i,prox(P_i))$. Then we select all its neighbors in a fixed range. Then we use this set of points to build the new line. To do so, we use the GEM, by providing adapted weights: 1 for the chosen points and 0 for the others. For numerical reason, we avoid to use 0 so we use $10^{-5}$ and $1-10^{-5}$. The result is a new mixture integrating this new line.

\begin{figure}[ht!]
  \centering
  \includegraphics[width=.49\textwidth]{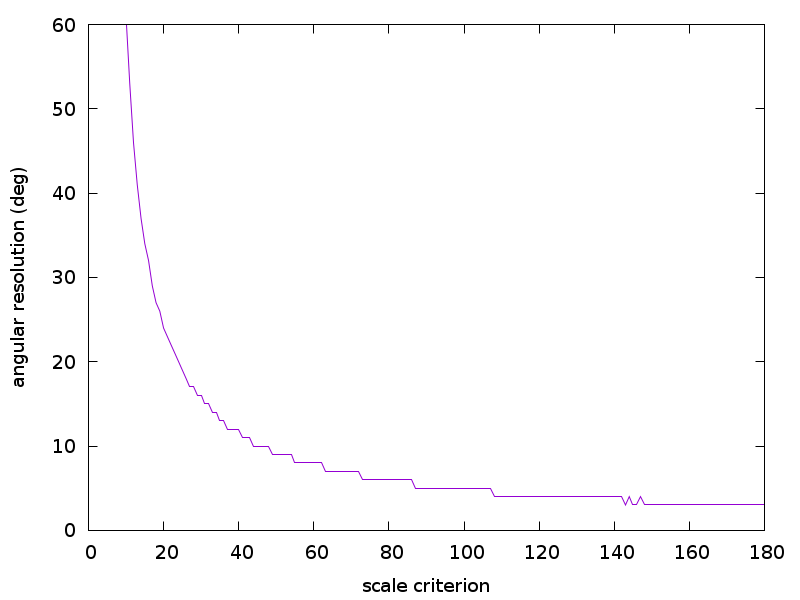}
  \hfill
  \includegraphics[width=.49\textwidth]{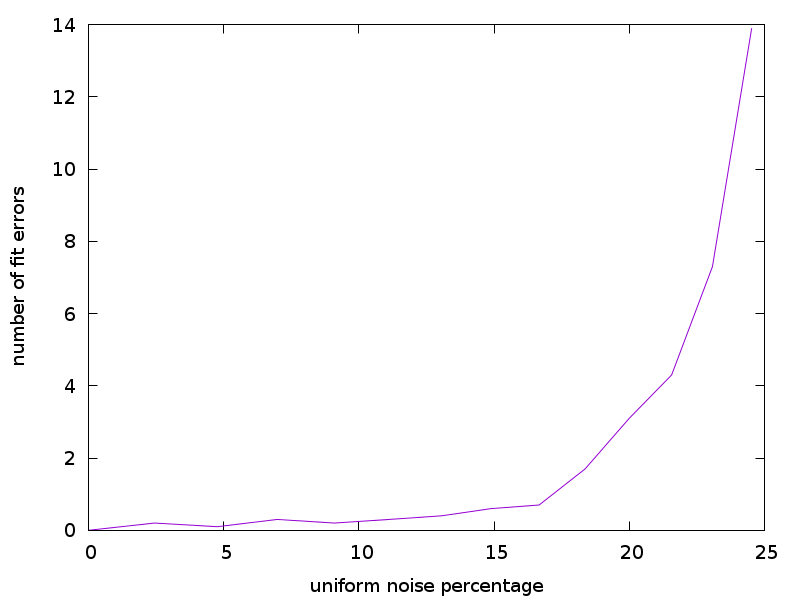}
  \caption{Angular resolution depending on scale criterion (left) and uniform noise resistance (right)}
  \label{fig:gem_props}
\end{figure}

To decide if a new line is necessary or not, we compare the standard deviation of all Gaussian error distributions to a scale criterion. This criterion allows one to choose the size of the distribution around the lines, regulating the dispersion of the data. It is defined as a fraction of the diagonal of the data space. It is also a criterion for choosing the angular resolution of the algorithm. As shown on Figure \ref{fig:gem_props} (on the left), the angular resolution depends on it. 

The MFit algorithm is naturally resistant to uniform noise. On the right part of the Figure \ref{fig:gem_props}, the uniform noise resistance is presented. We can see that up to 15\%, the noise is harmless but it becomes quickly intractable around 20\%. Indeed, as the percentage increases, the system tends to find alignments in the noise. 

On another hand, as MFit algorithm is purely additive, it happens frequently that, for numerical reasons, some lines are added in the mixture and stay in it even if they does not rely on any points. We call them degenerated lines. For performance reason, it is necessary to apply a small cut on the result to remove them. This cut has a linear complexity and is described in algorithm \ref{alg:degen}.

\begin{algorithm}[H]
  \For{$D_j \in D$}{ 
    $c=0$\;
    \For{$P_i \in P$}{
      \If{$D_j=prox(P_i)$}{$c=c+1$}
    }
    \If{$c<s$}{Remove($D_j$)}
  }
  \caption{bad fitted lines removal algorithm}
  \label{alg:degen}
\end{algorithm} 

\begin{figure}[ht!]
  \centering
  \includegraphics[width=.49\textwidth]{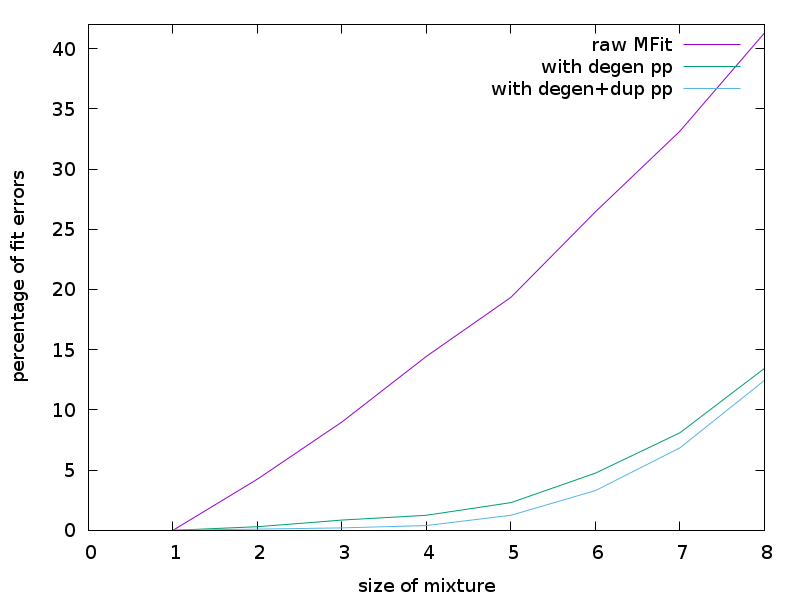}
  \caption{Performance of MFit algorithm on simulations with different post-processings.}
  \label{fig:gem_perf}
\end{figure}

To measure the global performance of the algorithm and the interest of the degenerated line post-processing, a simulation has been performed. Mixtures of non-colinear lines of different size are generated in 2 dimensions (2000 per size). On each of them, a fit is performed with the MFit algorithm. Figure \ref{fig:gem_perf} shows the obtained performance as the percentage of error on the size of the obtained mixture depending on its real size. We can see that without any post-processing (purple line), the performance is quickly degraded by the spurious lines. With degenerated line post-processing (green line), the performance is really better. A qualitative study showed that some lines are also duplicated. They can be suppressed by a specific post-processing but with a complexity of $O(J^2)$ with $J$ the number of component in the mixture. Figure \ref{fig:gem_perf} shows the effect of this second post-processing (blue line). The gain is constant and thus the performance is not increased a lot by its use. Thus, for applications, only the degenerated line post-processing is used.

\section{Applications}

All the algorithms described in this article have been implemented in a dedicated library called ``libgem''. It is an open-source software and can be downloaded on github at address {\em https://github.com/fredllr/libgem}.

It has been written in C in order to get the best possible performance. Bindings for other languages will be available soon. Command line programs allow to use main algorithms on csv data files. They are fully documented on the website. The code can also be interfaced with existing analysis framework. The API is pretty simple: a dataset must be filled with point coordinates and then the different algorithms can be called on it. The library also provides visualizations in both 2D and 3D, as shown on all the fit Figures of this article (except LRMM one, which are produced by R).

This implementation of MFit has been tested on the Harpo pre-processed data.

\begin{figure}[ht!]
  \centering
  \includegraphics[width=.40\textwidth]{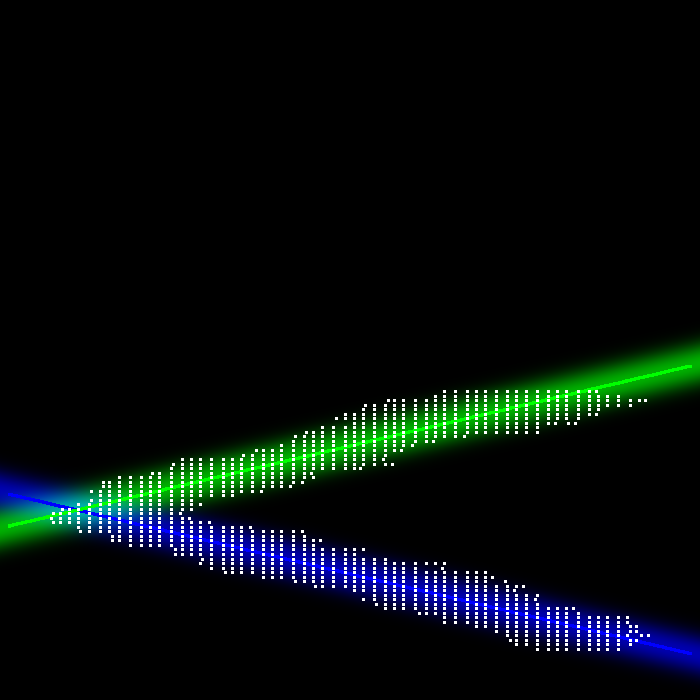}
  \hfill
  \includegraphics[width=.40\textwidth]{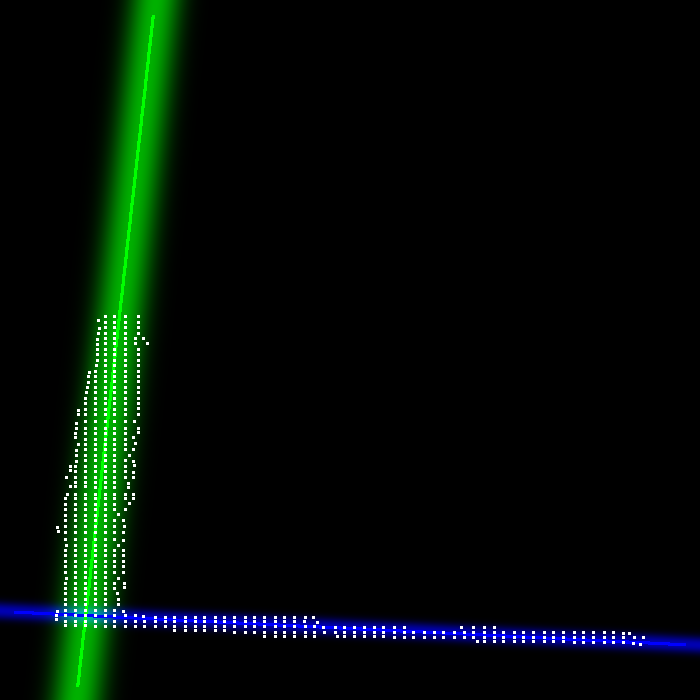}
  \caption{Harpo pair tracks in normal or vertical position, fitted with MFit.}
  \label{fig:harpo_gemfit_norm}
\end{figure} 

Figure \ref{fig:harpo_gemfit_norm} shows two cases of pair conversion (the dominant process at this energy). The algorithm detects correctly that there are two tracks. The width of each line depends on the data dispersion and the vertical line is not a problem anymore. The intersection of the two lines gives the interaction vertex, necessary for the reconstruction. 

\begin{figure}[ht!]
  \centering
  \includegraphics[width=.40\textwidth]{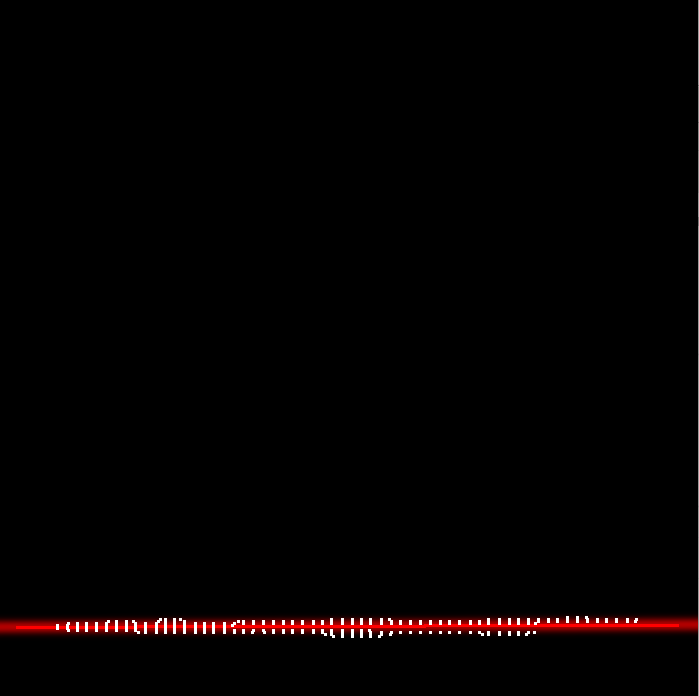}
  \hfill
  \includegraphics[width=.40\textwidth]{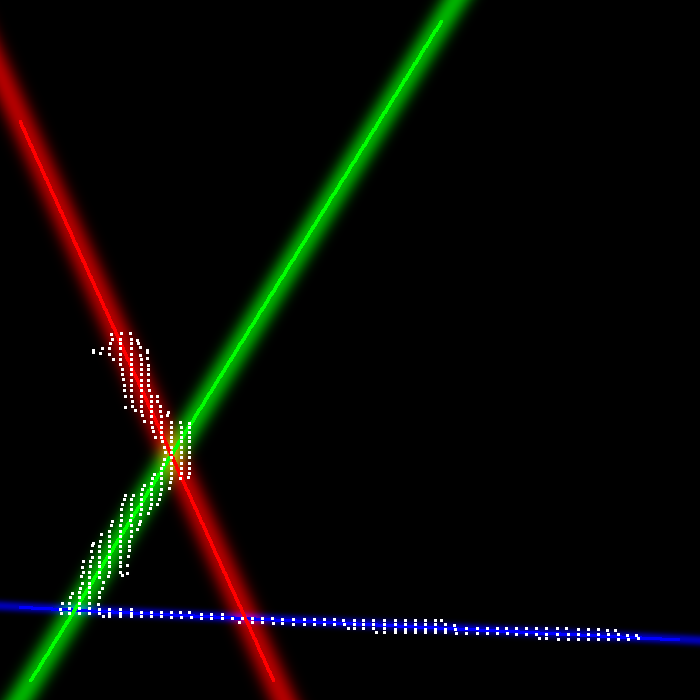}
  \caption{Fit of a Compton scattering and a pair creation with multiple scattering using MFit.}
  \label{fig:harpo_gemfit_diff}
\end{figure} 

Figure \ref{fig:harpo_gemfit_diff} shows two other cases. On the left, a unique track, probably a Compton scattering. On the right, a pair with so much scattering that the higher track has changed its direction and thus the algorithm added a line to adapt the fit to this situation. 

\begin{figure}[ht!]
  \centering
  \includegraphics[width=.40\textwidth]{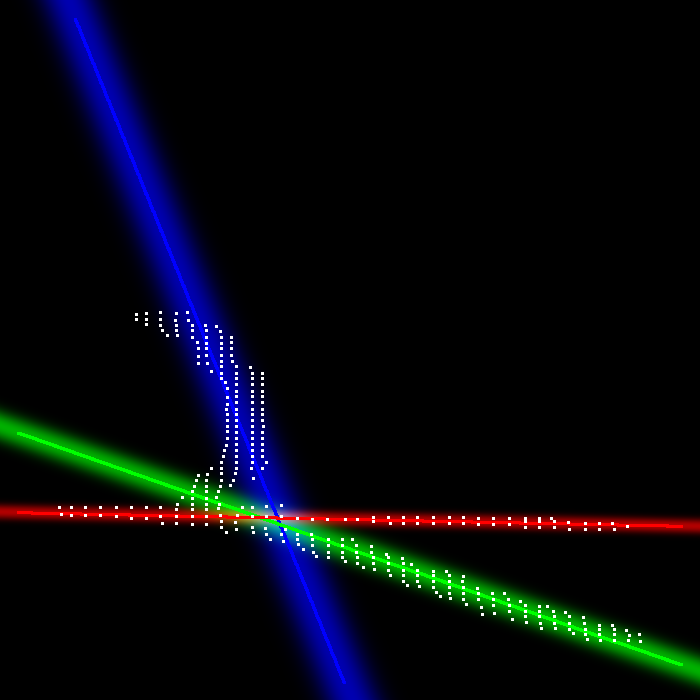}
  \hfill
  \includegraphics[width=.40\textwidth]{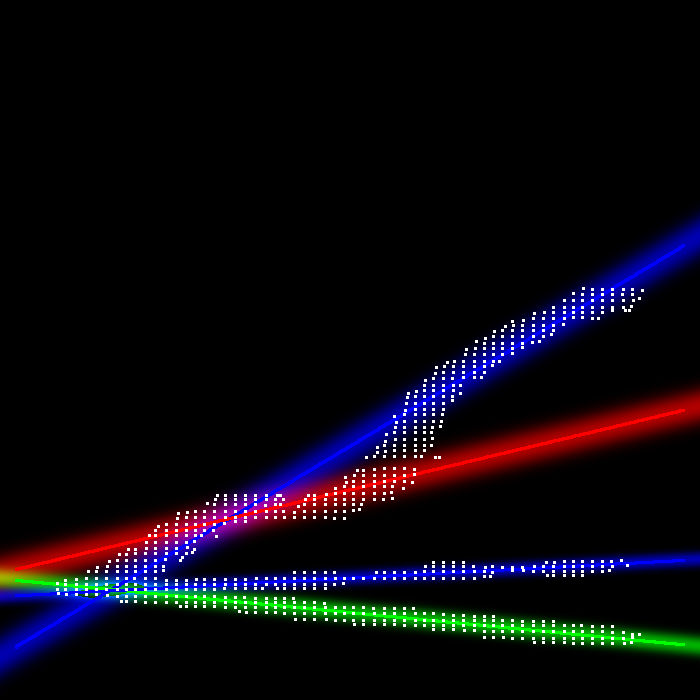}
  \caption{Triplets with multiple scatteting fitted with MFit.}
  \label{fig:harpo_gemfit_triple}
\end{figure} 

Figure \ref{fig:harpo_gemfit_triple} shows triplet's tracks. They can be distinguished from the previous case by the fact that the three-intersection points are very close and represent the vertex. On the right, again, the scattering has perturbed the line and the algorithm added a new line to cope with this variation.

The SiW-Ecal detector has been tested extensively at DESY during summer 2017. For MIP estimation, a lot of runs have been done without tungsten. In that case, the tracks are linear in a 3D space. Some pile-up is observed and two tracks are observed at the same time with a probability around 10\%. As energy deposit in silicon is proportional to material length, the angle of the incident tracks is important to estimate precisely the MIP (ionization minimum) value. The implementation, based on ``libgem'' is completely online, allowing to fit the tracks in real-time and display them.

\begin{figure}[h!]
  \centering
  \includegraphics[height=.45\textwidth]{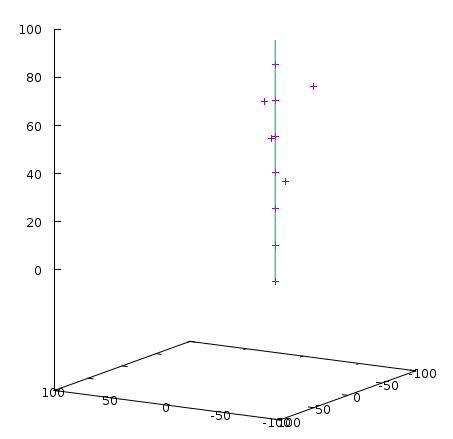}
  \hfill
  \includegraphics[height=.45\textwidth]{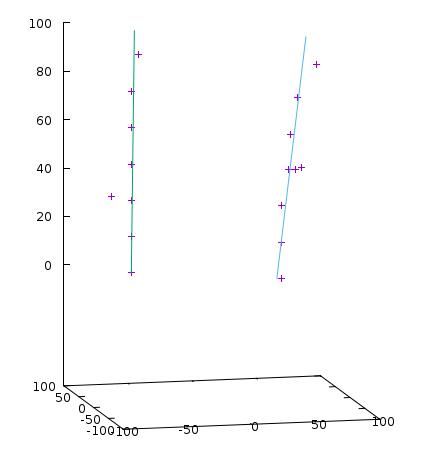}
  \caption{SiW-Ecal tracks with or without pile-up, fitted with MFit.}
  \label{fig:ecal_gem}
\end{figure}

Figure \ref{fig:ecal_gem} shows such a fit on 1 or 2 tracks. Tracks are slightly noised by showering. The fit matches the data and chooses the proper number of lines.

In such an online system, the performance of the algorithm has to be very good to cope with the data acquisition rate. As Mfit relies on $\Phi_{line}$ which is linear, the only delay is induced by the convergence loop itself which depends on the number of lines. In some complicated cases, the algorithm diverges, adding more and more lines without being able to find a satisfying fit. In those particular cases, the time of execution becomes out of scope of an online run. This is why it is necessary to add some protection, inside the algorithm, against these cases. In particular, it is useful to limit the number of added lines to some reasonable number. In the same way, we can limit the number of iterations and throw away the cases where the algorithm diverges.

\section{Extension to other shapes}

As mentioned before, we have managed that nothing in GEM algorithm is related to the linear shape of the fit. From the fact that all the geometric part is inside $\Phi_{line}$, we can infere a property of this algorithm.

\begin{prop}
  For any shape S, if an algorithm $\Phi_{S}$ exists, providing a weighted fit for this shape and an orthogonal distance to it, GEM can fit a mixture of multiple shapes S.
\end{prop}

Thus, GEM is a generic extension of EM algorithm to any shape for which a weighted fit is available. This is why we called it GEM for ``Generalized EM''. 

\begin{figure}[ht!]
\centering
\includegraphics[width=.49\textwidth]{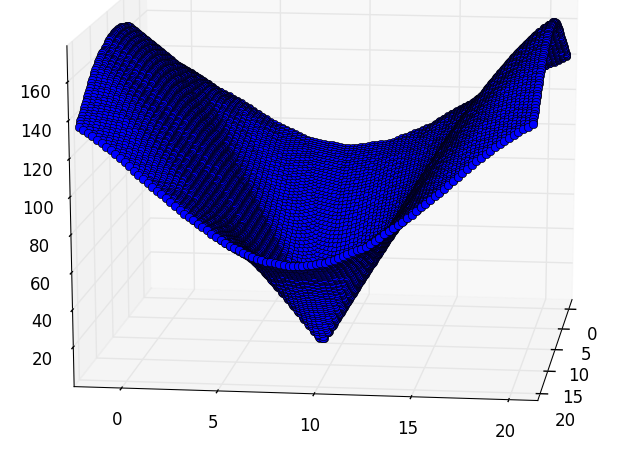}
\hfill
\includegraphics[width=.49\textwidth]{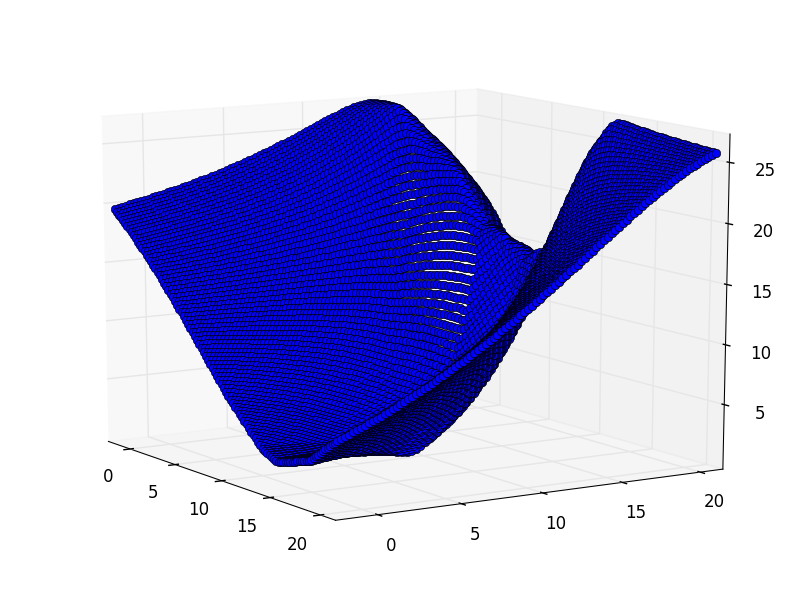}
\caption{Weighted error function of 20 points forming a circular track - Weighted error function of 7 points forming a quarter of a circular track.}
\label{fig:gradient}
\end{figure}

To demonstrate this availability, we propose to fit circular tracks. Indeed, circular tracks have a great interest in particle physics, as they correspond to the trajectory of charged particles in a uniform magnetic field. The problem of fitting a circle is inherently more complex than a linear fit because no analytical solution can be found. Thus an iterative algorithm has to be used to explore the space described by the weighted error function and to search for a minimum. The usual strategy for such a search is a line search strategy on the weighted error function defined by

\begin{equation}
\varepsilon(C)=\sum_{i=1}^n{\frac{\tau_i(\bar{d}-d(C,P_i))}{\sum_{j=1}^n{\tau_j}}},
\end{equation}

with $C$ denoting the expected center, the $P_i$ are the points to fit and the $\tau_i$ are their weights. The mean distance from the $P_i$ to $C$ is denoted $\bar{d}$. As shown on the left of Figure \ref{fig:gradient}, for a circular track, this function has a regular shape, a single minimum and a steep slope to reach it, making optimization easy. If the circle is incomplete, as shown on the right, the shape of the function is still regular and always includes a single minimum if at least three points are provided.

The traditional way to fit such function is to use a descent algorithm based on the gradient of the error function, such as the Newton-Raphson algorithm \cite{newtonbook}. The values of the coordinate of $C$ are evaluated iteratively using the partial derivative of the error function. The drawback of this algorithm is that the numerical evaluation of the partial derivatives is very time-consuming, thus, we propose a heuristic. Instead of calculating the descent direction with a derivative, we use the direction from the points to the previous version of the center.

\begin{figure}[ht!]
\centering
\includegraphics[width=.49\textwidth]{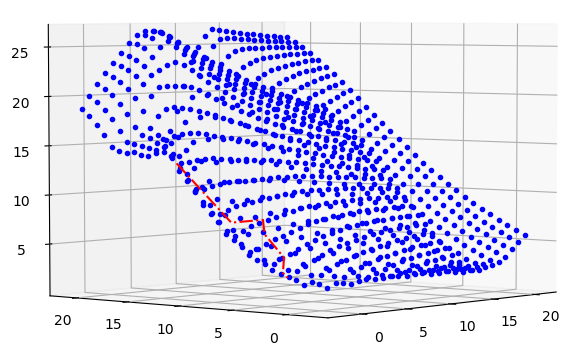}
\hfill
\includegraphics[width=.49\textwidth]{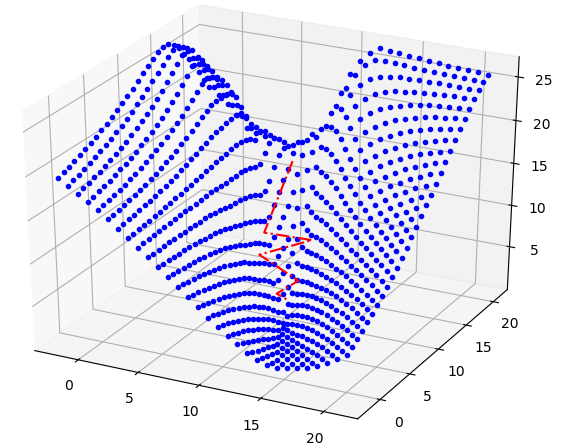}
\caption{Iterations of the center of the fitted center on seven points forming a quarter of a circular track, viewed from two angles.}
\label{fig:path}
\end{figure}

At the beginning, the center is initialized to the weighted barycenter of the points. Then, at each step, an error vector is computed for each point, directed to the expected center and proportional to the weighted error. Then a descent direction and amplitude are computed at the same time, as the vectorial sum of these individual error vectors. This gives a new center for the next iteration. The iterations are stopped once the position of the center is stable with respect to a parametrized threshold. Figure \ref{fig:path} shows the successive positions of the fit on the error function.

\SetKwInOut{Variables}{Variables}
\begin{algorithm}[H]
  \Variables{$C$ : center of the circle\\$r$ : radius of the circle\\$\overrightarrow{M_i}$ : move vectors\\$\varepsilon$ : array of distance between points and circle}
  $C=weighted\_barycenter(P,\tau)$\;
 \Repeat{center stable} {
   $\bar{d}=\frac{\sum_{i=1}^{n}{||\overrightarrow{P_iC}||}}{n}$\;
   \For{$i \in 1..n$} {
     $\displaystyle \overrightarrow{M_i}=\frac{\overrightarrow{P_iC}}{||\overrightarrow{P_iC}||}  \sum_{i=1}^{n}{\frac{\tau_i(\bar{d}-||\overrightarrow{P_iC}||)}{\sum_{i=1}^{n}{\tau_i}}}$\;
   }
   $C=C+\sum_{i=1}^{n}\overrightarrow{M_i}$\;
 }
 $r=\frac{\sum_{i=1}^{n}\tau_i \mid\mid\overrightarrow{CP_i}\mid\mid}{\sum_{i=1}^{n}\tau_i}$\;
  \For{$i \in 1..n$}{
    $\varepsilon_i=d(P_i,C)-r$\;
  }
  \Return $(C,r),\varepsilon$\;
  \caption{$\Phi_{circle}$ algorithm}
  \label{alg:phicirc}
\end{algorithm} 

\begin{figure}[ht!]
  \centering
  \includegraphics[width=.46\textwidth]{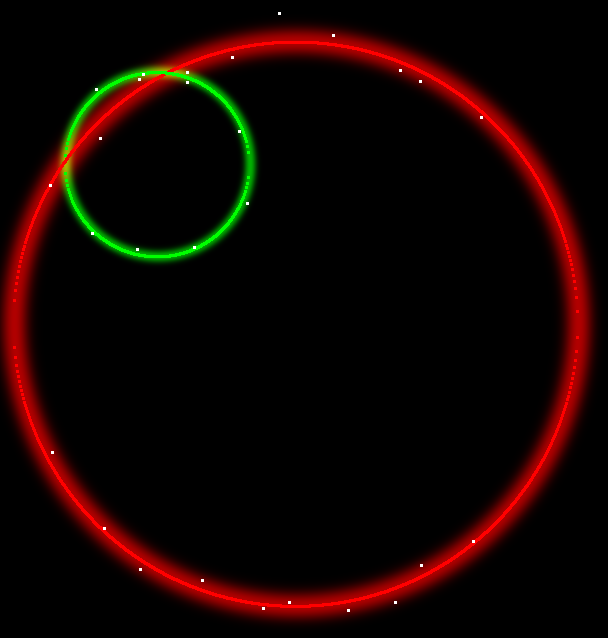}
  \hfill
  \includegraphics[width=.49\textwidth]{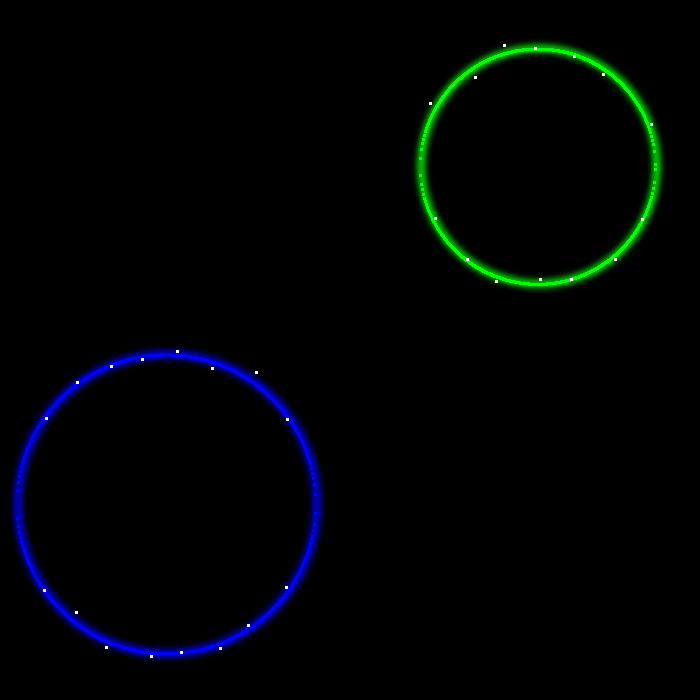}
  \caption{Adaptive circular fit on two circles, separated or entangled, with the MFit and $\Phi_{circle}$ algorithms.}
  \label{fig:mcfit}
\end{figure}

Of course, the algorithm MFit is naturally compatible with the new version and can fit an arbitrary number of circles at the same time without any modification. Figure \ref{fig:mcfit} shows such a double fit with mixed or separate circle tracks.

\section{Combined fits}

Now that we have two kinds of shape for our fit, we would like to combine them. Indeed, in MFit, there are no difference between $\Phi_{line}$ and $\Phi_{circle}$ and we can imagine to use them at the same time. The only difficulty is to find the good mixture to fit optimally the data. 

In this section, we propose an algorithm, named UFit, which tries different mixtures of different shapes and evaluates them with the previous score system $Score=\sum_{i=1}^{n} d(P_i,prox(P_i))$. UFit explores sequentially the mixtures [L] (only a line), [C] (a circle), [LL] (2 lines), [LC], [CL], [CC], [LLL]... It is necessary to explore both [LC] and [CL] because MFit is not commutative (the worst fitted point is very different if the first fit is a line or a circle). The algorithm stops as soon as all shapes of the mix are scaled.

UFit gives nice results if the number of shapes is less than 8. If we go above to this number, the complexity in $2^k$ becomes intractable. In order to test the algorithm, we made a simulation in Geant4: 1 GeV electrons are sent on a little cube of tungsten of 50 $\mu$m under a magnetic field of 0.3T. The electrons are showering in the tungsten and we get nice shower with various numbers of linear and circular tracks. The following plots present the different cases.
\begin{figure}[ht!]
  \centering
  \includegraphics[width=.3\textwidth]{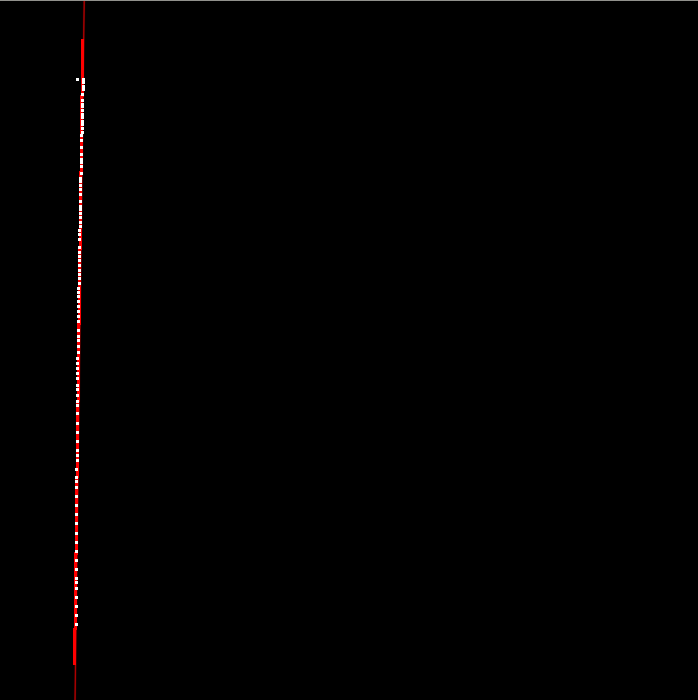}
  \includegraphics[width=.3\textwidth]{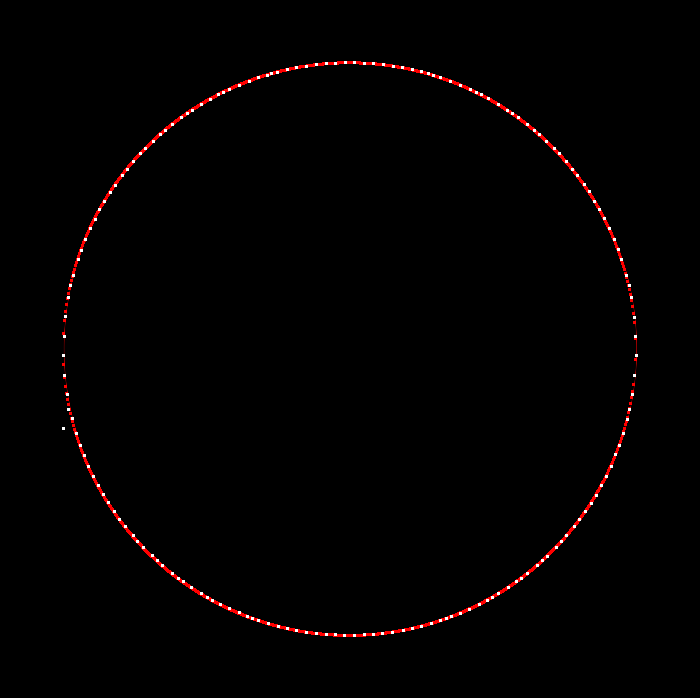}
  \includegraphics[width=.3\textwidth]{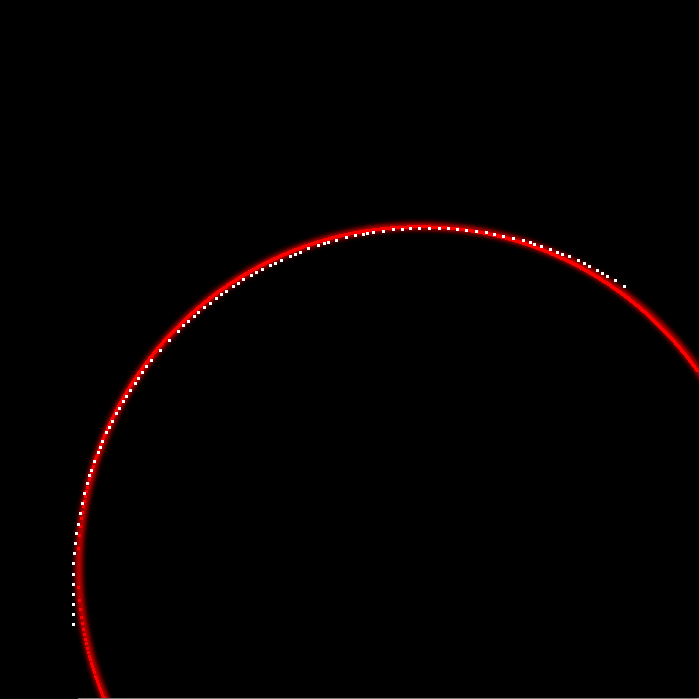}
  \caption{Ufit, on 1 track, discriminates correctly the base shapes.}
  \label{fig:ufit_1t}
\end{figure} 

On Figure \ref{fig:ufit_1t}, we can see that UFit discriminates correctly the two kinds of shapes if they are isolated, even if the circle is incomplete. For these very simple cases, only two calls to GEM algorithm are performed leading to a good performance.

\begin{figure}[ht!]
  \centering
  \includegraphics[width=.3\textwidth]{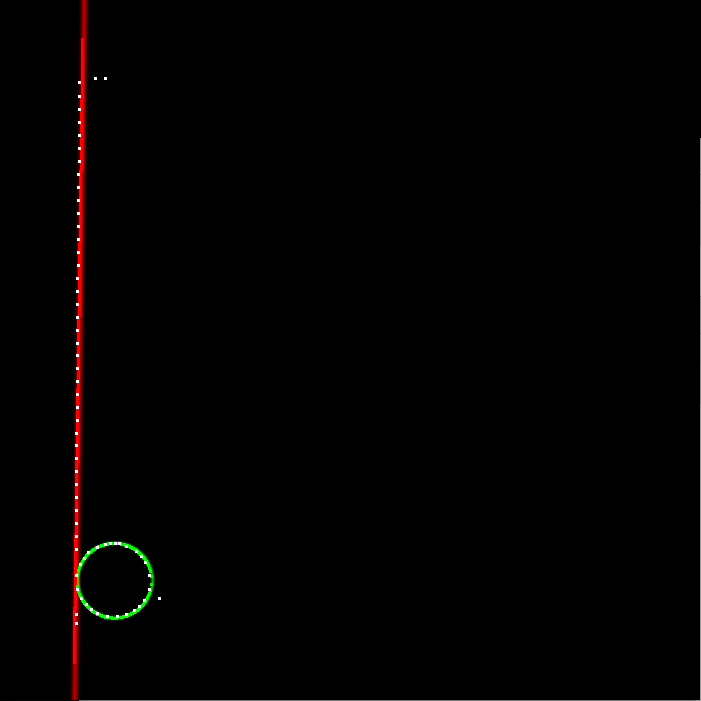}
  \includegraphics[width=.3\textwidth]{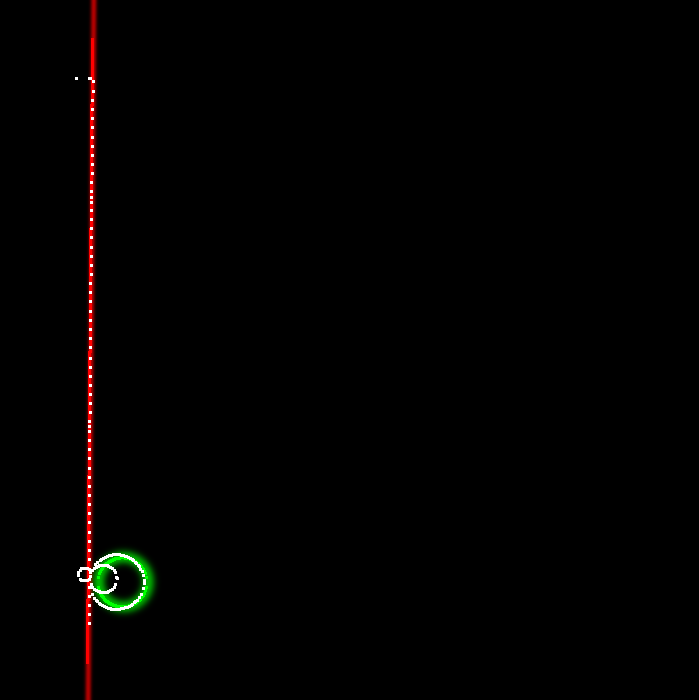}
  \includegraphics[width=.3\textwidth]{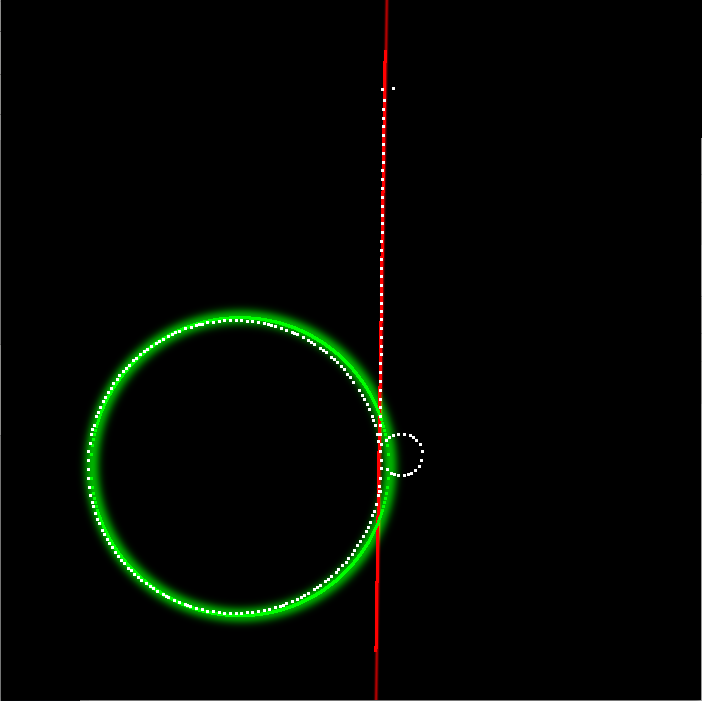}
  \caption{Ufit can fit two mixed tracks. The smaller structure are ignored if their typical size is comparable to Gaussian error standard deviation.}
  \label{fig:ufit_2t}
\end{figure} 

When two tracks are mixed, like on Figure \ref{fig:ufit_2t}, the algorithm extracts the two shapes. We can see that if very small structures are mixed, they are logically ignored if their typical size is the width of the Gaussian error. 

\begin{figure}[ht!]
  \centering
  \includegraphics[width=.3\textwidth]{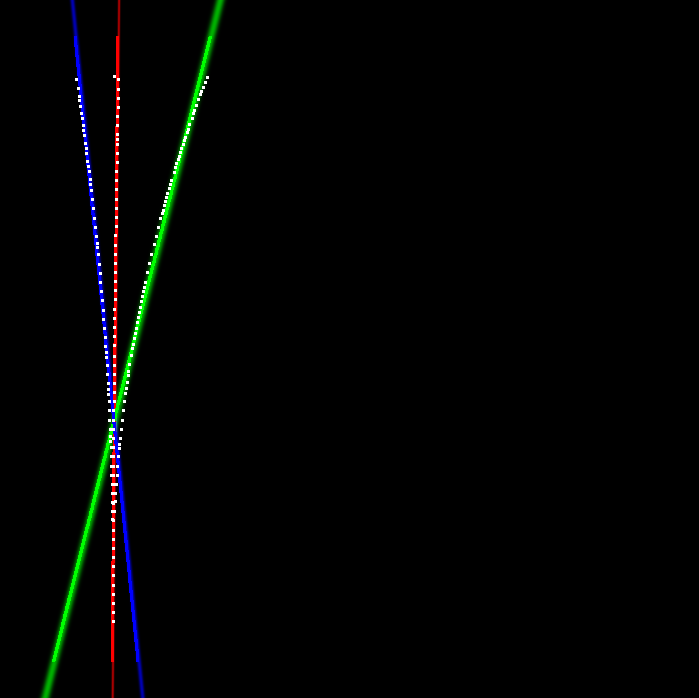}
  \includegraphics[width=.3\textwidth]{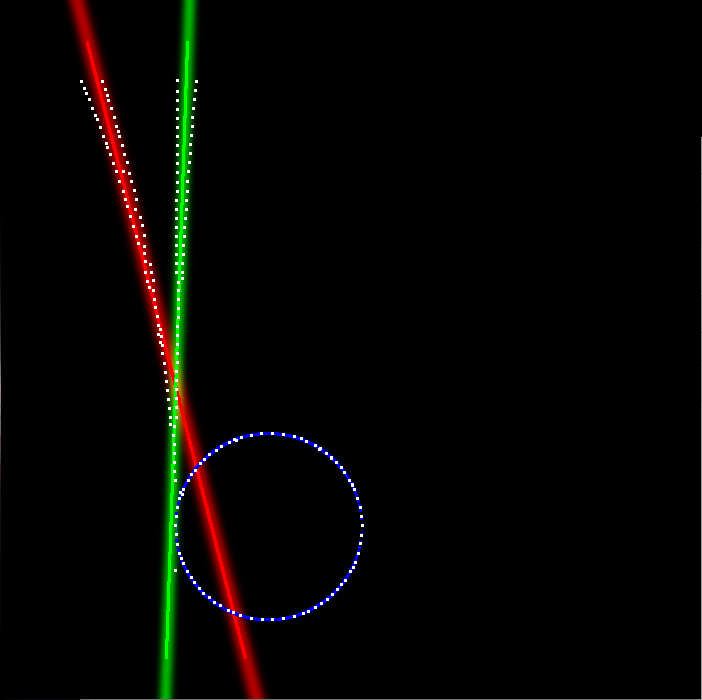}
  \includegraphics[width=.3\textwidth]{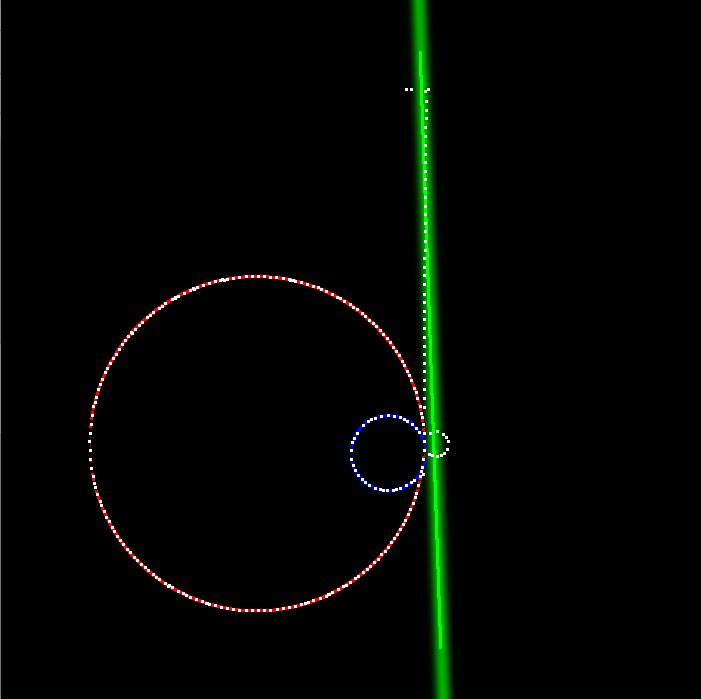}
  \caption{Ufit can fit three mixed tracks with 14 calls to GEM algorithm.}
  \label{fig:ufit_3t}
\end{figure} 

UFit algorithm gives its best when the mix is composed of 3 to 5 tracks. The number of calls to GEM stays reasonable (14 calls for 3 tracks, 30 for 4 and 62 for 5) and the discrimination power of the algorithm is good. As seen on Figure \ref{fig:ufit_3t}, the main tracks are properly fitted. On the middle plot, we can see that the tracks are showering a little but not enough to be considered as split lines (because their distance is behind the scale criterion). 

\begin{figure}[ht!]
  \centering
  \includegraphics[width=.3\textwidth]{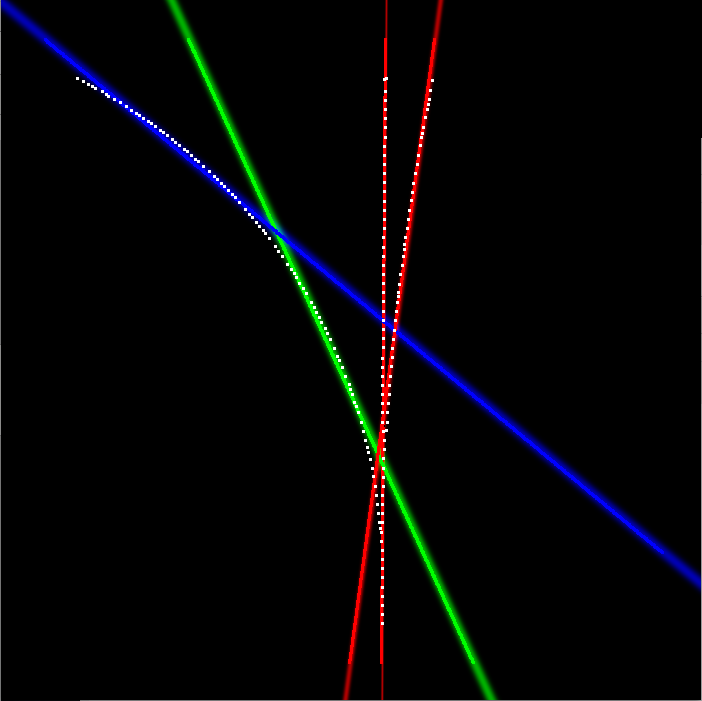}
  \includegraphics[width=.3\textwidth]{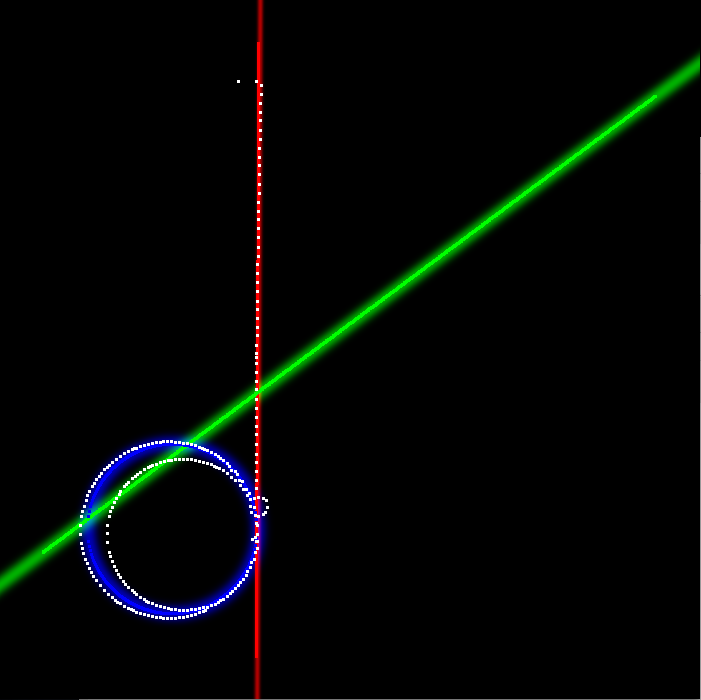}
  \includegraphics[width=.3\textwidth]{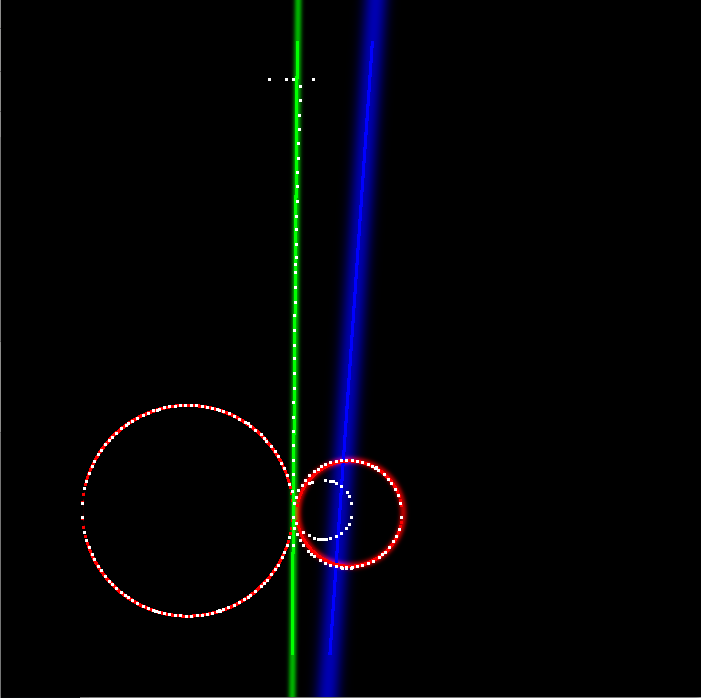}
  \caption{In some bad cases, Ufit adds a line instead of a circle due to its incremental nature.}
  \label{fig:ufit_badC}
\end{figure} 

In some case, UFit fails to see a circle because a lot of points are already involved in another shape. In that case, the convexity of the remaining points does not provide enough information to fit a circle. Figure \ref{fig:ufit_badC} shows some of these cases. On the left, the circle is badly fitted because the green line takes all lower points. In the middle, the blue circle prevent for fitting the inner circle. It is the same on the right. 

\begin{figure}[ht!]
  \centering
  \includegraphics[width=.3\textwidth]{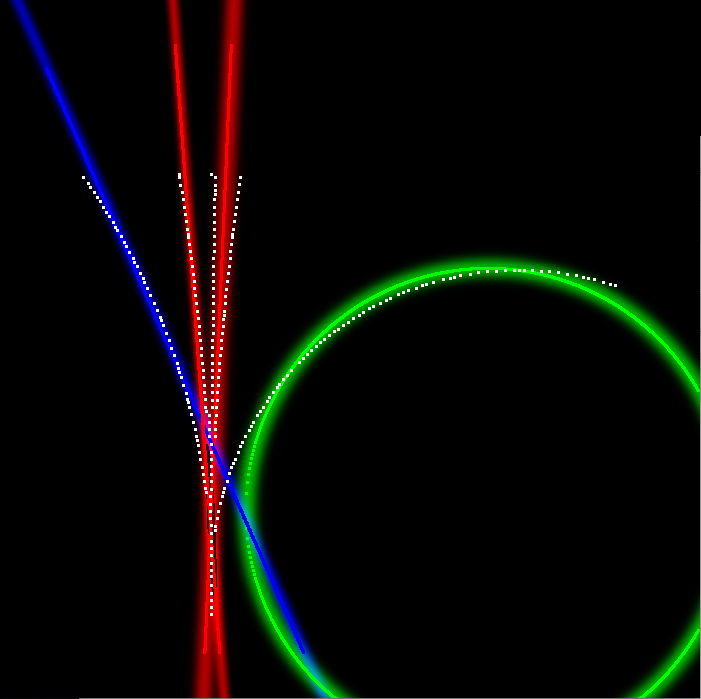}
  \includegraphics[width=.3\textwidth]{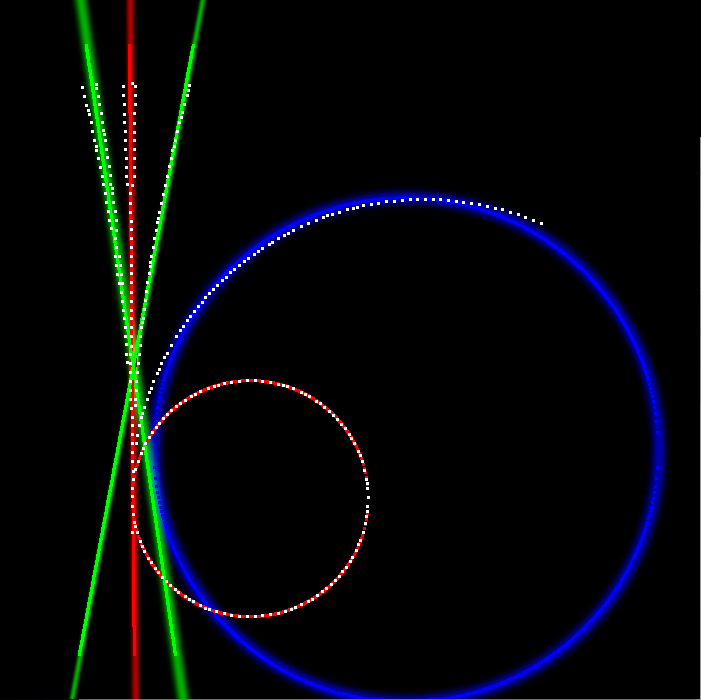}
  \includegraphics[width=.3\textwidth]{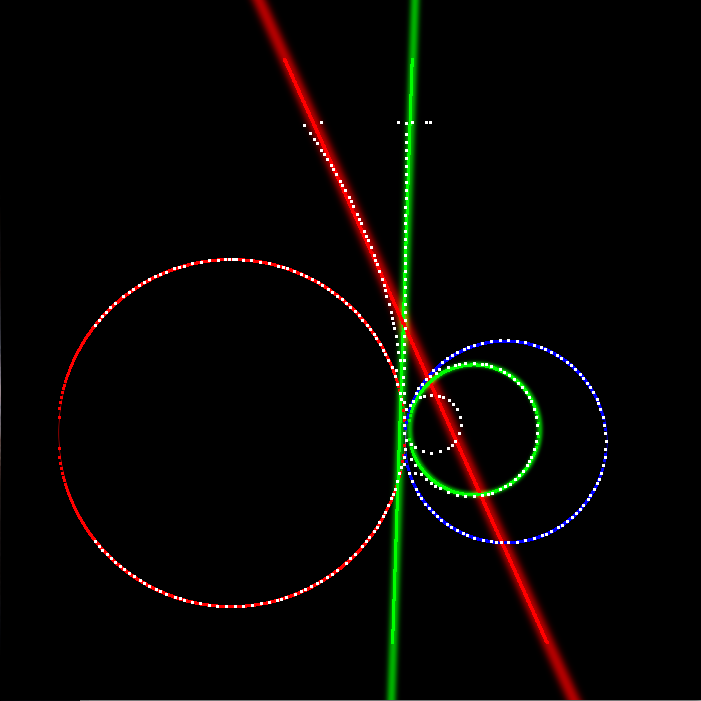}
  \caption{Ufit can handle up to 5 tracks with a performance which is compatible with online treatment.}
  \label{fig:ufit_5t}
\end{figure} 

UFit can discriminate easily 5 entangled tracks at the same time. Figure \ref{fig:ufit_5t} shows a nice mix correctly handled. We can see that in some cases (green circle on left and blue in the middle), the fitted circle is a little smaller than the track. This happens for the same reasons, the vertical part of the circle is involved in a line track and distort slightly the fit.

\begin{figure}[ht!]
  \centering
  \includegraphics[width=.4\textwidth]{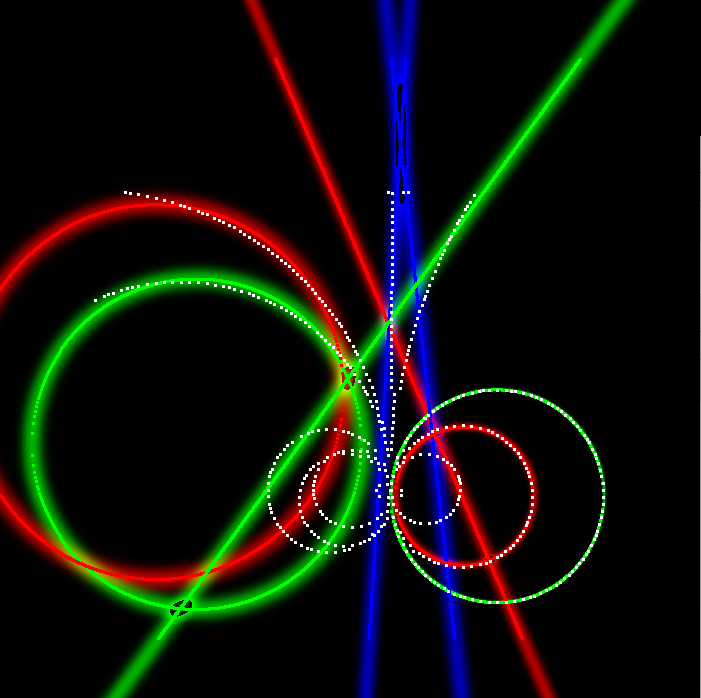}
  \caption{If too many tracks are entangled (more than 5), Ufit misses some shapes and its calculation time becomes too high.}
  \label{fig:ufit_complex}
\end{figure} 

Finally, if the number of tracks is too high, the algorithm misses some of them because, the points are inside the error distribution and are not considered as candidates for new shapes. Another problem comes from the computation time which becomes too high for online use.

\section{Conclusion}

The EM algorithm is a great tool for extracting information from raw data in various scientific fields. It can also be used in particle trajectography. The algorithm proposed in \cite{turner2000} presents some isotropy problems and dimension limitation. By using orthogonal distances, we can propose an alternative estimator solving the anisotropy but also an alternative model giving the extension to arbitrary dimension. The MFit algorithm solves two other problems of EM algorithm: touchy initialization and adaptive size of the mixture. We have seen that the problem is extendable to other shapes, with circular tracks as an example and that the shapes are even mixable in the UFit algorithm.

This work can receive many further developments. The linear fit can be enriched by considering only segments instead of full lines. If the intersection points can be identified (by their weights), complex shapes composed of these segments could be fitted. In particular, this should help to fit electromagnetic or hadronic showers as they appear in calorimeters. 

The circular extension can fit spheres in 3D. This could be extended to fit ellipsoid by estimating the direction and the eccentricity parameters. This is interesting to quickly identify a shower versus a hadronic jet. 

Another extension for allowing GEM and derivate algorithms to work on bigger mixtures could be based on a partition of space. The data space could be split in many parts, on which GEM would be applied. The results would then be aggregated to get the global mixture. This could be very helpful to provide detailed tracks of very complicated collision induced by pile-up in high-luminosity detectors. 

Finally, the helix shape could be studied, representing a charged particle in a uniform magnetic field with a direction which is not in the longitudinal axis of the detector. The combined use of $\Phi_{line}$ and $\Phi_{circle}$ allows to estimate the parameters of this shape, but at that time, there is no way to distinguish entangled helices. A meta-algorithm could probably lead to weighted helix fit and then to multiple helices fit.

\newpage
\section*{Acknowledgements}

The author warmly thanks Philippe Gros for bringing this problem to his attention and helping him with Harpo data and simulation. The author would also like to thank Marie-Laure Martin-Magniette for her precious help on mixture models and EM algorithms. 

\bibliographystyle{unsrt} 
\bibliography{gem}

\end{document}